\documentclass[proceedings]{stacs}
\stacsheading{2010}{47-58}{Nancy, France}
\firstpageno{47}

\usepackage{amsmath}
\usepackage{amssymb}
\usepackage{mathtools}
\usepackage{pifont}
\usepackage{paralist}
\usepackage{phdalgo}
\usepackage{ifthen}
\usepackage{hypergraph}
\usepackage{bold-extra}
\usepackage{pgf}
\usepackage{tikz}
\usetikzlibrary{calc}
\usetikzlibrary{arrows}
\usetikzlibrary{shapes}
\colorlet{myblue}{gray!50!black}
\colorlet{myred}{gray!50!black}
\colorlet{mygreen}{gray!50!black}
\DeclareSymbolFontAlphabet{\mathbb}{AMSb}
\DeclareMathAlphabet{\mathbbb}{U}{bbold}{m}{n}
\DeclareSymbolFont{stmry}{U}{stmry}{m}{n}
\DeclareMathSymbol\mapsfromchar\mathrel{stmry}{"5B}

\newcommand{\href}[2]{#2}
\newcommand{\mpplus}{\oplus}
\newcommand{\mptimes}{\otimes}

\newcommand{\maxplus}{\mathbb{R}_{\max}}
\newcommand{\defin}{:=}
\newcommand{\mpzero}{\mathbbb{0}}
\newcommand{\mpone}{\mathbbb{1}}

\newcommand{\mponevect}{\pmb{\mpone}}
\newcommand{\ie}{\textit{i.e.}}

\newcommand{\mpsum}{\mathop{\bigoplus}}
\newcommand{\vect}[1]{\boldsymbol #1}
\newcommand{\PP}{\mathcal{P}}
\newcommand{\CC}{\mathcal{C}}
\newcommand{\KK}{\mathcal{K}}
\newcommand{\HH}{\mathcal{H}}
\newcommand{\DD}{\mathcal{D}}

\newcommand{\supp}{\operatorname{\mathsf{supp}}}
\newcommand{\argmax}{\arg\max}
\newcommand{\tangent}{\mathcal{T}}

\newcommand{\size}{\mathsf{size}}
\newcommand{\card}[1]{\left\lvert #1 \right\rvert}

\newcommand{\upperbound}[2]{U(#1,#2)}
\newcommand{\mychoose}[2]{\binom{#1}{#2}}
\newcommand{\mpzeroone}{\{\,\mpzero,\mpone\,\}}
\newcommand{\mponealt}{\mpone}

\newcommand{\etal}{\textit{et al.}}

\newcommand{\R}{\mathbb{R}}

\newcommand{\rbar}{\R\cup\{-\infty\}}

\newcommand{\reachable}{\rightsquigarrow}
\newcommand{\reach}{\reachable}

\newcommand{\reacheq}{\equiv}
\newcommand{\reachleq}{\preceq}

\newcommand{\ComputeExtremal}{\Call{ComputeExtreme}{}}

\newcommand{\scc}{\textsc{Scc}}

\algrenewcommand\alglinenumber[1]{\scriptsize #1:}
\renewcommand{\lineref}[1]{{\scriptsize\ref{#1}}}
\renewcommand{\baryx}{\vect{x}_1}
\renewcommand{\baryy}{\vect{x}_2}
\renewcommand{\baryz}{\vect{x}_3}

\newboolean{include_appendix}
\setboolean{include_appendix}{false}
{\bfseries}{\itshape}

\begin{document}
\title{The tropical double description method}
\author[ref1]{Xavier {A}llamigeon}{Xavier {A}llamigeon}
\address[ref1]{Direction du Budget, 4\`eme sous-direction, Bureau des transports, Paris, France}
\thanks{This work was performed when the first author was with EADS Innovation Works, SE/IA -- Suresnes, France and CEA, LIST MeASI -- Gif-sur-Yvette, France}
\thanks{This work was partially supported by the Arpege programme of the French National Agency of Research (ANR), project ``ASOPT'', number ANR-08-SEGI-005 and
by the Digiteo project DIM08 ``PASO'' number 3389}
\author[ref2]{{S}t{\'e}phane {G}aubert}{{S}t{\'e}phane {G}aubert}
\address[ref2]{INRIA Saclay and CMAP, Ecole Polytechnique, France}
\author[ref3]{{\'E}ric {G}oubault}{{\'E}ric {G}oubault}
\address[ref3]{CEA, LIST MeASI -- Gif-sur-Yvette, France}
\email{firstname.lastname@{polytechnique.org,inria.fr,cea.fr}}
\begin{abstract}
We develop a tropical analogue of the classical double description
method allowing one to compute an internal
representation (in terms of vertices) of a polyhedron defined
externally (by inequalities).  The heart of the tropical algorithm is
a characterization of the extreme points of a polyhedron in terms of a
system of constraints which define it. We show that checking the
extremality of a point reduces to checking whether there is only one
minimal strongly connected component in an hypergraph. The latter
problem can be solved in almost linear time, which allows us to
eliminate quickly redundant generators. We report extensive tests
(including benchmarks from an application to static analysis) showing
that the method outperforms experimentally the previous ones by orders
of magnitude. The present tools also lead to worst case bounds which
improve the ones provided by previous methods.
\end{abstract}

\maketitle

\section*{Introduction}\label{sec:intro}
Tropical polyhedra are the analogues of convex polyhedra in tropical algebra. The latter deals with structures like the max-plus semiring $\maxplus$ (also called \emph{max-plus algebra}), which is the set $\mathbb{R} \cup \{-\infty\}$, equipped with the addition $x \mpplus y \defin \max(x,y)$ and the multiplication $x \mptimes y \defin x + y$. 

The study of the analogues of convex sets in tropical or max-plus algebra is an active research topic, and has been treated under various guises. It arose in the work of Zimmerman~\cite{zimmerman77}, following a way opened by Vorobyev~\cite{vorobyev67}, motivated by optimization theory.
Max-plus cones 
were studied by Cuninghame-Green~\cite{CG}. Their theory was independently developed by Litvinov, Maslov and Shpiz~\cite{litvinov00} (see also~\cite{maslov92}) with motivations from variations calculus and asymptotic
analysis, and by Cohen, Gaubert, and Quadrat~\cite{cgq02} 
who initiated a ``geometric approach'' of discrete event systems~\cite{ccggq99}, further developed in~\cite{katz05,loiseau}. Other motivations
arise from abstract convexity, see the book by Singer~\cite{ACA},
and also the work of Briec and Horvath~\cite{BriecHorvath04}. 
The field has attracted recently more attention after the work of Develin and Sturmfels~\cite{DS}, who pointed out connections with tropical geometry, leading to several works 
by Joswig, Yu, and the same authors~\cite{joswig04,DevelinYu,JSY07,joswig-2008}.

A tropical polyhedron can be represented in two different ways, either internally, in terms of extreme points and rays, 
or externally, in terms of linear inequalities (see Sect.~\ref{sec:definition} for details). As in the classical case, passing from the external description of a polyhedron to its internal description is a fundamental computational issue. This is the object of the present paper.

Butkovi\v{c} and Hegedus~\cite{butkovicH} gave an algorithm to compute the generators of a tropical polyhedral cone 
described by linear inequalities. Gaubert gave a similar one and derived the equivalence between the internal and external representations~\cite[Ch.~III]{gaubert92a} (see~\cite{GK09b} for a recent discussion). Both algorithms rely on a successive elimination of inequalities, but have the inconvenience of squaring at each step the number of candidate generators, unless an elimination technique is used, as in the Maxplus toolbox of {\sc Scilab}~\cite{toolbox}. Joswig developed a different approach, implemented in {\sc Polymake}~\cite{polymake}, in which a tropical polytope is represented as a polyhedral complex~\cite{DS,joswig-2008}. 


The present work grew out from two applications: to discrete event systems~\cite{katz05,loiseau}, and to software verification by static analysis~\cite{AGG08}.
In these applications, passing
from the external to the internal representation is a central
difficulty. A further motivation originates from mean payoff games~\cite{AGGut09}. These motivations are reviewed in Section~\ref{sec-motiv}.


\medskip\noindent{\em Contributions.}
We develop a new algorithm which computes the extreme elements of tropical polyhedra. It is based on a successive elimination of inequalities, and 
a result (Th.~\ref{th:ddm}) allowing one, given a polyhedron $\PP$ and a tropical halfspace $\HH$, to construct
a list of candidates for the generators of $\PP \cap \HH$.
The key ingredient is a combinatorial characterization of the extreme generators
of a polyhedron defined externally (Th.~\ref{th:criterion2} and~\ref{th:minimality}): we reduce the verification of the extremality of a candidate to the existence of a strongly connected component reachable from any other in a directed hypergraph. 
We include a complexity analysis and experimental results (Sect.~\ref{sec:ddm}),
showing that the new algorithm outperforms the earlier ones, allowing us to solve instances which were previously by far inaccessible. Our result
also leads to worst case bounds improving the ones of previously known algorithms. 

\section{Definitions: tropical polyhedra and polyhedral cones}\label{sec:definition}
The neutral elements for the addition $\mpplus$ and multiplication $\mptimes$, \ie{}, the zero and the unit, will be denoted by $\mpzero\defin-\infty$ and $\mpone\defin 0$, respectively.
The tropical analogues of the operations on vectors and matrices are
defined naturally. The elements of $\maxplus^d$, 
the $d$th fold Cartesian product of $\maxplus$,
will be thought of as vectors, and denoted by bold symbols, 
like $\vect x=(\vect{x}_1,\ldots, \vect{x}_d)$.

A \emph{tropical halfspace} is a set of the vectors $\vect{x} = (\vect{x}_i)\in \maxplus^d$ verifying an inequality constraint of the form 
\[
\max_{1\leq i\leq d} a_i +\vect{x}_i \leq \max_{1\leq i\leq d} b_i + \vect{x}_i, \quad a_i, b_i \in \maxplus.
\]
A \emph{tropical polyhedral cone} is 
defined as the intersection of $n$ halfspaces. It can be equivalently written as the set of the solutions of a system of inequality constraints $A \vect{x} \leq B \vect{x}$.
Here, $A=(a_{ij})$ and $B=(b_{ij})$ are $n \times d$ matrices with entries in $\maxplus$, concatenation denotes the matrix product (with the laws of $\maxplus$), and $\leq$ denotes
the standard partial ordering of vectors. For sake of readability, tropical polyhedral cones will be simply referred to as \emph{polyhedral cones} or \emph{cones}.


Tropical polyhedral cones are known to be generated by their extreme rays~\cite{GK06a,GK,BSS}. 
Recall that a {\em ray} 
is the set of scalar multiples of a non-zero vector $\vect u$. It is {\em extreme} in a cone $\CC$ if $\vect u\in \CC$ and if $\vect u=\vect v \mpplus \vect w$ with $\vect v,\vect w\in \CC$ implies that $\vect u=\vect v$ or $\vect u=\vect w$. A finite set $G = (\vect{g}^i)_{i \in I}$ of vectors is said to {\em generate} a polyhedral cone $\CC$ if each
$\vect{g}^i$ belongs to $\CC$, and if every vector $\vect{x}$ of $\CC$
can be written as a \emph{tropical linear combination} $\mpsum_i \lambda_i \vect{g}^i$ of the vectors
of $G$ (with $\lambda_i \in \maxplus$). Note that in tropical linear combinations, the requirement that $\lambda_i$ be nonnegative is omitted. Indeed, $\mpzero =-\infty\leq \lambda$ holds for all scalar $\lambda\in \maxplus$.

The tropical analogue of the Minkowski theorem~\cite{GK,BSS} shows in particular that every generating set of a cone that is minimal for inclusion is obtained by selecting precisely one (non-zero) element in each extreme ray. 


A tropical polyhedron of $\maxplus^d$ is the affine analogue of a tropical polyhedral cone. It is defined by a system of inequalities of the form $A\vect x\oplus \vect{c}\leq B\vect x\oplus \vect{d}$. It can be also expressed as the set of the tropical affine combinations of its generators. The latter are of the form $\mpsum_{i \in I} \lambda_i  \vect{v}^i \,\, \mpplus \,\, \mpsum_{j \in J} \mu_j \vect{r}^j$, where the $(\vect{v}^i)_{i \in I}$ are the extreme points, the $(\vect{r}^j)_{j \in J}$ a set formed by one element of each extreme ray, and $\mpsum_i \lambda_i = \mpone$.
It is known~\cite{cgq02,GK} that every tropical polyhedron of $\maxplus^d$ can be represented by a tropical polyhedral cone of $\maxplus^{d+1}$ thanks to an analogue of the homogenization method used in the classical case (see~\cite[Sect.~1.5]{ziegler98}). Then, the extreme rays of the cone are in one-to-one correspondence with the extreme generators of the polyhedron. That is why, in the present paper, we will only state the main results for cones, leaving to the reader the derivation of the affine analogues, along the lines of~\cite{GK}.
In the sequel, we will illustrate our results on the polyhedral cone $\CC$ given in Fig.~\ref{fig:tropical_poly}, defined by the system in the right side. 
The left side is a representation of $\CC$ in barycentric coordinates: each element $(\vect{x}_1,\vect{x}_2,\vect{x}_3)$ is represented as a barycenter with weights $(e^{\vect{x}_1}, e^{\vect{x}_2}, e^{\vect{x}_3})$ of the three vertices of the outermost triangle. Then two
elements of a same ray are represented by the same point. 
The cone $\CC$ is depicted in solid gray (the black border is included), and is generated by the extreme elements $\vect{g}^0 = (\mpzero,0,\mpzero)$, $\vect{g}^1 = (-2,1,0)$, $\vect{g}^2 = (2,2,0)$, and $\vect{g}^3 = (0,\mpzero,0)$.

\begin{figure}
\begin{center}
\begin{minipage}[c]{0.6\textwidth}
\vskip-3ex
\begin{center}
\begin{tikzpicture}
[scale=0.65,>=triangle 45
,vtx/.style={mygreen},
ray/.style={myred}]
\equilateral{7}{100};

\barycenter{g1}{\expo{0}}{0}{\expo{0}};
\barycenter{g12}{\expo{0}}{\expo{0}}{\expo{0}};
\barycenter{g2}{\expo{2}}{\expo{2}}{\expo{0}};
\barycenter{g3}{0}{\expo{0}}{0};
\barycenter{g4}{\expo{-2}}{\expo{1}}{\expo{0}};
\barycenter{g41}{\expo{0}}{\expo{1}}{\expo{0}};

\filldraw[lightgray,draw=black,very thick] (g1) -- (g12) -- (g2) -- (g3) -- (g4) -- (g41) -- cycle;
\filldraw[vtx] (g4) circle (0.75ex) node[above left=1pt] {$\vect{g}^1$};
\filldraw[vtx] (g2) circle (0.75ex) node[left=1pt] {$\vect{g}^2$};
\filldraw[vtx] (g1) circle (0.75ex) node[above right=1pt] {$\vect{g}^3$};
\filldraw[ray] (g3) circle (0.75ex) node[above right=1pt] {$\vect{g}^0$};

\node at (9.5,4) {
$
\left\{
\begin{aligned}
\vect{x}_3 & \leq \vect{x}_1+2 \\
\vect{x}_1 & \leq \max(\vect{x}_2,\vect{x}_3) \\
\vect{x}_1 & \leq \vect{x}_3+2 \\
\vect{x}_3 & \leq \max(\vect{x}_1,\vect{x}_2-1)
\end{aligned}
\right.$};
\end{tikzpicture}
\end{center}

\caption{A tropical polyhedral cone in $\maxplus^3$}\label{fig:tropical_poly}
\end{minipage}
\hfill
\begin{minipage}{0.39\textwidth}
\vskip-3ex
\hskip-3ex
\begin{tikzpicture}[join=round]
\begin{scope}[scale=0.5]
\tikzstyle{axe}=[lightgray,>=triangle 45];
\draw[arrows=->,axe](0,-5.714)--(0,5.714);
\draw[arrows=-,axe](.436,.19)--(-.582,-.254);
\draw[arrows=-,axe](-.873,.095)--(1.164,-.127);
\draw[thick](0,0)--(-2.182,-.952);
\draw[arrows=-,axe](-.582,-.254)--(-2.182,-.952);
\filldraw[fill=gray,fill opacity=0.3,thick,draw=black](0,0)--(2.182,3.333)--(-2.182,-.952)--(-2.182,-5.714)--(0,-4.762)--cycle;
\draw[thick](0,0)--(4.364,-.476);
\draw[arrows=-,axe](1.164,-.127)--(4.364,-.476);
\filldraw[fill=gray,fill opacity=0.3,thick,draw=black](0,0)--(0,-4.762)--(4.364,-5.238)--(4.364,-.476)--(2.182,3.333)--cycle;
\filldraw[thick,fill opacity=0.3,fill=gray,draw=black](0,0)--(2.182,3.333)--(0,4.762)--cycle;
\filldraw[fill=gray,fill opacity=0.3,thick,draw=black](4.364,-.476)--(2.182,3.333)--(2.182,-1.429)--(2.182,-6.19)--(4.364,-5.238)--cycle;
\draw[arrows=->,axe](4.364,-.476)--(5.237,-.571);
\filldraw[fill=gray,fill opacity=0.3,thick,draw=black](-2.182,-.952)--(-2.182,-5.714)--(2.182,-6.19)--(2.182,-1.429)--(2.182,3.333)--cycle;
\draw[arrows=->,axe](-2.182,-.952)--(-2.837,-1.238);
\path (-2.837,-1.238) node[gray,left] {$\mathtt{len\_src}$}
                   (5.237,-.571) node[gray,above] {$\mathtt{len\_dst}$}
                   (0,5.714) node[gray,left] {$\mathtt{n}$};
\end{scope}
\end{tikzpicture}
\caption{\texttt{memcpy} invariant}\label{fig:memcpy_invariant}
\end{minipage}
\end{center}
\end{figure}

\section{Motivations from static analysis, discrete event systems, and mean payoff games}\label{sec-motiv}
Tropical polyhedra have been recently involved in static analysis by abstract interpretation~\cite{AGG08}. It has been shown that they allow to automatically compute complex invariants involving the operators $\min$ and $\max$ which hold over the variables of a program. Such invariants are disjunctive, while most existing techniques in abstract interpretation are only able to express conjunctions of affine constraints, see in particular~\cite{CousotCousot77-1,CousotHalbwachs78-POPL,MineAst01}. 

For instance, tropical polyhedra can handle notorious problems in verification of memory manipulations. Consider the well-known memory string manipulating function \texttt{memcpy} in C. A call to $\mathtt{memcpy(dst,src,n)}$ copies exactly the first $\mathtt{n}$ characters of the string buffer $\mathtt{src}$ to $\mathtt{dst}$. In program verification, precise invariants over the length of the strings are needed to ensure the absence of string buffer overflows. Recall that the length of a string is defined as the position of the first null character in the string. To precisely analyze the function \texttt{memcpy}, two cases have to be distinguished: 
\begin{enumerate}[(i)]
\item either $\mathtt{n}$ is strictly smaller than the source length $\mathtt{len\_src}$, so that only non-null characters are copied into $\mathtt{dst}$, hence $\mathtt{len\_dst} \geq \mathtt{n}$,
\item or $\mathtt{n} \geq \mathtt{len\_src}$ and the null terminal character of $\mathtt{src}$ will be copied into $\mathtt{dst}$, thus $\mathtt{len\_dst} = \mathtt{len\_src}$.
\end{enumerate}
Thanks to tropical polyhedra, the invariant $\min(\mathtt{len\_src},\mathtt{n}) = \min(\mathtt{len\_dst},\mathtt{n})$, or equivalently $\max(-\mathtt{len\_src},-\mathtt{n}) = \max(-\mathtt{len\_dst},-\mathtt{n})$, can be automatically inferred. It is the \emph{exact} encoding of the disjunction of the cases (i) and (ii). The invariant is represented by the non-convex set of $\mathbb{R}^3$ depicted in Figure~\ref{fig:memcpy_invariant}.
In the application to static analysis, the performance of the algorithm computing the extreme elements of tropical polyhedra plays a crucial role in the scalability of the analyzer (see~\cite{AGG08} for further details).

A second motivation arises from the ``geometric approach'' of max-plus linear
discrete event systems~\cite{ccggq99}, in which the computation of feedbacks
ensuring that the state of the system meets a prescribed
constraint (for instance that certain waiting times remain bounded)
reduces~\cite{katz05} to computing the greatest fixed
point of an order preserving map on the set of tropical
polyhedra. Similar computations arise when solving
dual observability problems~\cite{loiseau}.
Again, the effective handling of these polyhedra turns
out to be the bottleneck.

A third motivation arises from the study of mean payoff
combinatorial games. In particular, it is shown in~\cite{AGGut09}
that checking whether a given initial state of a mean payoff game is
winning is equivalent to finding a vector in an associated
tropical polyhedral cone (with a prescribed finite coordinate).
This polyhedron consists
of the super-fixed points of the dynamic programming operator
(potentials), which certify that the game is winning.

\section{Characterizing extremality from inequality constraints}\label{sec:extremality_criterion}

\subsection{Preliminaries on extremality}

The following lemma, which is a variation on the proof of Th.~3.1 of~\cite{GK} and on Th.~14 of~\cite{BSS}, 
shows that extremality can be expressed as a minimality property:
\begin{proposition}\label{prop:minimality}
Given a polyhedral cone $\CC \subset \maxplus^d$, $\vect{g}$ is extreme if and only if there exists $1 \leq t \leq d$ such that $\vect{g}$ is a minimal element of the set $\{\, \vect{x} \in \CC \mid \vect{x}_t = \vect{g}_t \,\}$, \ie{} $\vect g\in \CC$ and for each $\vect{x} \in \CC$, 
$\vect{x} \leq \vect{g} \text{ and } \vect{x}_t = \vect{g}_t \text{ implies } \vect{x} = \vect{g}.$
In that case, $\vect{g}$ is said to be \emph{extreme of type $t$}.
\end{proposition}

%
%
%
%

In Fig.~\ref{fig:type}, the light gray area represents the set of the elements $(\vect{x}_1,\vect{x}_2,\vect{x}_3)$ of $\maxplus^3$ such that $(\vect{x}_1,\vect{x}_2,\vect{x}_3) \leq \vect{g}^2$ implies $\vect{x}_1 < \vect{g}^2_1$. It clearly contains the whole cone except $\vect{g}^2$, which shows that $\vect{g}^2$ is extreme of type $1$.

A {\em tropical segment} is the set of the tropical 
linear 
 combinations of two points. Using the fact
that a tropical segment joining two points of a polyhedral cone $\CC$
yields a continuous path included in $\CC$, one can check that $\vect g$ is extreme of type $t$ in $\CC$ if and only if 
there is a neighborhood $N$ of $\vect{g}$ such that $\vect{g}$ is minimal in $\{\, \vect{x} \in \CC \cap N \mid \vect{x}_t = \vect{g}_t \,\}$. Thus, extremality is a local property.

Finally, the extremality of an element $\vect{g}$ in a cone $\CC$ can be equivalently established by considering the vector formed by its non-$\mpzero$ coordinates. 
Formally, let $\supp(\vect{x}) \defin \{i\mid \vect{x}_i\neq\mpzero\}$ for any $\vect{x} \in \maxplus^d$.
Then $\vect{g}$ is extreme in $\CC$ if and only if it is extreme in $\{ \vect{x} \in \CC \mid \supp(\vect{x}) \subset \supp(\vect{g}) \}$. This allows to assume that $\supp(\vect{g}) =\{1,\ldots,d\}$ without loss of generality.

\subsection{Expressing extremality using the tangent cone} 

For now, the polyhedral cone $\CC$ is supposed to be defined by a system $A \vect{x} \leq B \vect{x}$ of $n$ inequalities. 
%

Consider an element $\vect{g}$ of the cone $\CC$,
which we assume, from the previous discussion,
to satisfy $\supp(\vect g)=\{\,1,\ldots,d\,\}$.
In this context, the \emph{tangent cone} of $\CC$ at $\vect{g}$ is defined as the tropical polyhedral cone $\tangent( \vect{g}, \CC)$ of $\maxplus^d$ given by the system of inequalities
\begin{equation}
\max_{i \in \argmax(A_k \vect{g})} \vect{x}_i \leq \max_{j \in \argmax(B_k \vect{g})} \vect{x}_j \qquad \text{for all }k \text{ such that }A_k \vect{g} = B_k \vect{g}, \label{eq:tangent_constraint}
\end{equation}
where for each row vector $\vect{c} \in \maxplus^{1 \times d}$, $\argmax(\vect{c} \vect{g})$ is defined as the argument of the maximum $\vect{c} \vect{g} = \max_{1 \leq i \leq d} (\vect{c}_i + \vect{g}_i)$, and where $A_k$ and $B_k$ denote the $k$th rows of $A$ and $B$, respectively.

The tangent cone $\tangent(\vect{g},\CC)$ provides a local description of the cone $\CC$ around $\vect{g}$:
\begin{proposition}\label{prop:tangent_cone}
There exists a neighborhood $N$ of $\vect{g}$ such that for all $\vect{x} \in N$, $\vect{x}$ belongs to $\CC$ if and only if it is an element of $\vect{g} + \tangent(\vect{g},\CC)$.
\end{proposition}

%
%
\begin{figure}[t]
\begin{center}
\begin{minipage}[b]{0.45\textwidth}
\vspace{0ex}
\vskip-3ex
\begin{center}
\begin{tikzpicture}
[scale=0.65,>=triangle 45
,vtx/.style={mygreen},
ray/.style={myred}]
\equilateral{7}{90};

\barycenter{g1}{\expo{0}}{0}{\expo{0}};
\barycenter{g12}{\expo{0}}{\expo{0}}{\expo{0}};
\barycenter{g2}{\expo{2}}{\expo{2}}{\expo{0}};
\barycenter{g3}{0}{\expo{0}}{0};
\barycenter{g4}{\expo{-2}}{\expo{1}}{\expo{0}};
\barycenter{g41}{\expo{0}}{\expo{1}}{\expo{0}};

\barycenter{g2y}{exp(1)}{0}{exp(0)};
\barycenter{g2z}{exp(1)}{exp(1)}{0};

\filldraw[lightgray,draw=black,very thick] (g1) -- (g12) -- (g2) -- (g3) -- (g4) -- (g41) -- cycle;

\filldraw[gray!40,draw=none,fill opacity=0.3] (g2) -- (g2z) -- (y) -- (z) -- (g2y) -- cycle;

\node[fill=black,circle,inner sep=0pt,minimum size=1.25ex] at (g2) [label=left:$\vect{g}^2$] {};
\end{tikzpicture}
\end{center}
\vskip-2ex
\caption{Extremality of $\vect{g}^2$}\label{fig:type}
\end{minipage}
\hfill
\begin{minipage}[b]{0.45\textwidth}
\vspace{0ex}
\vskip-3ex
\begin{center}
\begin{tikzpicture}
[scale=0.65,>=triangle 45
,vtx/.style={mygreen},
ray/.style={myred}]
\equilateral{7}{90};

\barycenter{g1}{\expo{0}}{0}{\expo{0}};
\barycenter{g12}{\expo{0}}{\expo{0}}{\expo{0}};
\barycenter{g2}{\expo{2}}{\expo{2}}{\expo{0}};
\barycenter{g3}{0}{\expo{0}}{0};
\barycenter{g4}{\expo{-2}}{\expo{1}}{\expo{0}};
\barycenter{g41}{\expo{0}}{\expo{1}}{\expo{0}};

\barycenter{g2y}{exp(1)}{0}{exp(0)};
\barycenter{g2z}{exp(1)}{exp(1)}{0};

\filldraw[lightgray,draw=black,very thick] (g1) -- (g12) -- (g2) -- (g3) -- (g4) -- (g41) -- cycle;

\filldraw[gray!40,draw=none,fill opacity=0.3] (z) -- (g2) -- (y);
\draw[gray, dotted, very thick] (z) -- (g2) -- (y) -- (z);

\node[circle,fill=black,inner sep=0pt,minimum size=1.25ex] at (g2) [label=left:$\vect{g}^2$] {};
\end{tikzpicture}
\end{center}
\vskip-2ex
\caption{The set $\vect{g}^2 + \tangent(\vect{g}^2,\CC)$}\label{fig:tangent1}
\end{minipage}
\end{center}
\end{figure}

As an illustration, Fig.~\ref{fig:tangent1} depicts the set $\vect{g}^2 + \tangent(\vect{g}^2,\CC)$ (in semi-transparent light gray) when $\CC$ is the cone given in Fig.~\ref{fig:tropical_poly}. Both clearly coincide in the neighborhood of $\vect{g}^2$.

Since extremality is a local property, it can be equivalently
characterized in terms of the tangent cone. Let $\mponevect$ be the element of $\maxplus^d$ whose all coordinates are equal to $\mpone$.
\begin{proposition}\label{prop:unit_extremality}
The element $\vect{g}$
is extreme in $\CC$ iff the vector $\mponevect$ is extreme in 
$\tangent(\vect{g}, \CC)$.
\end{proposition}
The problem is now reduced to the characterization of the extremality of the vector $\mponevect$ in a \emph{$\mpzeroone$-cone}, \ie{} a polyhedral cone defined by a system of the form $C \vect{x} \leq D \vect{x}$ where $C, D \in \mpzeroone^{n \times d}$. The following proposition states that only $\mpzeroone$-vectors, \ie{} elements of the tropical regular cube $\mpzeroone^d$, have to be considered:
%
\begin{proposition}\label{prop:criterion1}
Let $\DD \subset \maxplus^d$ be a $\mpzeroone$-cone. 
Then $\mponevect$ is extreme of type $t$ if and only if it is the unique element $\vect{x}$ of $\DD \cap \mpzeroone^d$ satisfying $\vect{x}_t = \mponealt$.
\end{proposition}

The following criterion of extremality is a direct consequence of Prop.~\ref{prop:unit_extremality} and~\ref{prop:criterion1}:
\begin{theorem}\label{th:criterion2}
Let $\CC \subset \maxplus^d$ be a polyhedral cone. Then $\vect{g} \in \CC$ is extreme of type $t$ if and only if the vector $\mponevect$ is the unique $\mpzeroone$-element of the tangent cone $\tangent(\vect{g},\CC)$ whose $t$-th coordinate is $\mponealt$.
\end{theorem}

Figure~\ref{fig:tangent2} shows that in our running example, the $\mpzeroone$-elements of $\tangent(\vect{g}^2, \CC)$ distinct from $\mponevect$ (in squares) all satisfy $\vect{x}_1 = \mpzero$.
Naturally, testing, by exploration, whether the set of $2^{d-1}$ $\mpzeroone$-elements $\vect{x}$ verifying $\vect{x}_t = \mponealt$ belonging to $\tangent(\vect{g},\CC)$ 
consists only of $\mponevect$ 
does not have an acceptable complexity. Instead, the approach of the next section will rely on the equivalent formulation of the criterion of Th.~\ref{th:criterion2}:
\begin{equation}
\forall l \in \{\, 1, \dots, d\, \}, \; \bigl[\forall \vect{x} \in \tangent(\vect{g},\CC) \cap \mpzeroone^d,\; \vect{x}_l = \mpzero \implies \vect{x}_t = \mpzero \bigr]\label{eq:criterion3}.
\end{equation}

\begin{figure}
\begin{minipage}[b]{0.5\textwidth}
\vskip-3ex
\begin{center}
\begin{tikzpicture}
[scale=0.65,>=triangle 45
,vtx/.style={mygreen},
ray/.style={myred},
lbl/.style={}]
\equilateral{7}{90};

\barycenter{unit}{\expo{0}}{\expo{0}}{\expo{0}};
\filldraw[lightgray,draw=black,very thick] (z) -- (unit) -- (y) -- cycle;

\node[circle,fill=black,inner sep=0pt,minimum size=1.25ex] at (unit) [label=below left:$\mponevect$] {};
\node[rectangle,fill=myblue,inner sep=0pt,minimum size=1.15ex] at (y) [label={[lbl]left:$(\mpzero, \mponealt, \mpzero)$}] {};
\node[rectangle,fill=myblue,inner sep=0pt,minimum size=1.15ex] at (z) [label={[lbl]right:$(\mpzero, \mpzero, \mponealt)$}] {};
\node[rectangle,fill=myblue,inner sep=0pt,minimum size=1.15ex] at (yz) [label={[lbl]right:$(\mpzero, \mponealt, \mponealt)$}] {};
\end{tikzpicture}
\end{center}
\vskip-2ex
\caption{The $\mpzeroone$-elements of $\tangent(\vect{g}^2,\CC)$}\label{fig:tangent2}
\end{minipage}
\hfill
\begin{minipage}[b]{0.45\textwidth}
\vskip-3ex
\begin{center}
\begin{tikzpicture}[>=stealth',scale=0.8, vertex/.style={circle,draw=black,very thick,minimum size=2ex}, hyperedge/.style={draw=black,thick}, simpleedge/.style={draw=black,thick}]
\node [vertex] (u) at (-2,-1.25) {$u$};
\node [vertex] (v) at (0,-0.25) {$v$};
\node [vertex] (w) at (0,-2)  {$w$};
\node [vertex] (x) at (3.5,-0.25) {$x$};
\node [vertex] (y) at (3.5,-2) {$y$};
\node [vertex] (t) at (2,-3.4) {$t$};

\path[->] (u) edge[simpleedge,out=90,in=-180] (v);
\node at (-1,-0.5) {$e_1$};
\path[->] (v) edge[simpleedge,out=-90,in=90] (w);
\node at (-0.5,-1.3) {$e_2$};
\path[->] (w) edge[simpleedge,out=-120,in=-60] (u);
\node at (-1.5,-2.5) {$e_3$};
\node at (1.75,-0.5) {$e_4$};
\node at (2.5,-2.75) {$e_5$};
\hyperedge{v}{v,w}{w}{0.35}{0.6}{y,x};
\hyperedge{y.west}{y.west,w.east}{w.east}{0.48}{0.5}{t};
\end{tikzpicture}
\end{center}
\vskip-2ex
\caption{A directed hypergraph}\label{fig:hypergraph}
\end{minipage}
\end{figure}

\subsection{Characterizing extremality with directed hypergraphs}

A \emph{directed hypergraph} is a couple $(N,E)$ such that each element of $E$ is of the form $(T,H)$ with $T, H \subset N$. 

The elements of $N$ and $E$ are respectively called \emph{nodes} and \emph{hyperedges}. Given a hyperedge $e = (T,H) \in E$, the sets $T$ and $H$ represent the \emph{tail} and the \emph{head} of $e$ respectively, and are also denoted by $T(e)$ and $H(e)$. Figure~\ref{fig:hypergraph} depicts an example of hypergraph whose nodes are $u, v, w, x, y, t$, and of hyperedges $e_1 =(\{ u\}, \{v\})$, $e_2 = (\{ v\}, \{w\})$, $e_3 = (\{w\}, \{u\})$, $e_4 = (\{ v,w \}, \{x,y\})$, and $e_5 = (\{ w,y \}, \{t\})$.

Reachability is extended from digraphs to directed hypergraphs by the following recursive definition: given $u,v \in N$, then $v$ is \emph{reachable from $u$ in $\HH$}, which is denoted $u \reachable_\HH v$, if one of the two conditions holds:
$u = v$, 
or there exists $e \in E$ such that $v \in H(e)$ and all the elements of $T(e)$ are reachable from $u$.
In our example, $t$ is reachable from $u$.

The size $\size(\HH)$ of a hypergraph $\HH = (N,E)$ is defined as $\card{N} + \sum_{e \in E} (\card{T(e)} + \card{H(e)})$. In the rest of the paper, directed hypergraphs will be simply referred to as hypergraphs.

We associate to the tangent cone $\tangent(\vect{g}, \CC)$ the hypergraph $\HH(\vect{g},\CC) = (N,E)$ defined by:
\[
N = \{\, 1,\dots,d\, \} \qquad E = \left\{\, (\argmax(B_k \vect{g}), \argmax(A_k \vect{g})) \mid A_k \vect{g} = B_k \vect{g}, \; 1 \leq k \leq n \,\right\}.
\]
The extremality criterion of Eq.~\eqref{eq:criterion3} suggests to evaluate, given an element of $\tangent(\vect{g},\CC) \cap \mpzeroone^d$, the effect of setting its $l$-th coordinate to the other coordinates. Suppose that it has been discovered that $\vect{x}_l = \mpzero$ implies $\vect{x}_{j_1} = \dots = \vect{x}_{j_n} = \mpzero$. For any hyperedge $e$ of $\HH(\vect{g},\CC)$ such that $T(e) \subset \{\, l,j_1,\dots,j_n \,\}$, $\vect{x}$ satisfies:
$
\max_{i \in H(e)} \vect{x}_i \leq \max_{j \in T(e)} \vect{x}_j = \mpzero
$,
so that $\vect{x}_i = \mpzero$ for all $i \in H(e)$. Thus, the propagation of the value $\mpzero$ from the $l$-th coordinate to other coordinates mimicks the inductive definition of the reachability relation from the node $l$ in $\HH(\vect{g},\CC)$:
\begin{proposition}\label{prop:reachability}
For all $l \in \{\,1,\dots,d\,\}$, the statement given between brackets in Eq.~\eqref{eq:criterion3} holds if and only if $t$ is reachable from $l$ in the hypergraph $\HH(\vect{g},\CC)$.
\end{proposition}
Hence, the extremality criterion can be restated thanks to some considerations on the strongly connected components of $\HH(\vect{g},\CC)$. The \emph{strongly connected components} (\scc{}s for short) of a hypergraph $\HH$ are the equivalence classes of the equivalence relation $\reacheq_\HH$, defined by $u \reacheq_\HH v$ if $u \reach_\HH v$ and $v \reach_\HH u$. They form a partition of the set of nodes of $\HH$. They can be partially ordered by the relation $\reachleq_\HH$, defined by $C_1 \reachleq_\HH C_2$ if $C_1$ and $C_2$ admit a representative $u$ and $v$ respectively such that $v \reach_\HH u$ (\emph{beware of the order of $v$ and $u$ in $v \reach_\HH u$}). Then Prop.~\ref{prop:reachability} and Th.~\ref{th:criterion2} imply the following statement:
\begin{theorem}\label{th:minimality}
Let $\CC \subset \maxplus^d$ be a polyhedral cone, and $\vect{g} \in \CC$. Then $\vect{g}$ is extreme if and only if the set of the \scc{}s of the hypergraph $\HH(\vect{g},\CC)$, partially ordered by $\reachleq_{\HH(\vect{g},\CC)}$, admits a least element. 
\end{theorem}
This theorem is reminiscent of a classical result,
showing that a point of a polyhedron defined by inequalities
is extreme if and only if the family of gradients of active inequalities
at this point is of full rank. Here, the hypergraph encodes precisely the
subdifferentials (set of generalized gradients) of the active inequalities
but a major difference is that the rank
condition must be replaced by the above minimality condition, which is
essentially stronger. Indeed, using this theorem, it is shown in~\cite{AGK09} that an important class of tropical polyhedra has fewer extreme rays than its classical analogue. 

An algorithm due to Gallo \etal{}~\cite{GalloDAM93} shows that one can compute the set of nodes that are reachable from a given node in linear time in an hypergraph. 
The following result shows that one can in fact compute
the minimal \scc{}s  with almost the same complexity.
The algorithm is included in the extended version of the present paper~\cite{AGG09}.
Although it shows
some analogy with the classical Tarjan algorithm, 
the hypergraph case differs critically from the graph case
in that one cannot compute all the \scc{}s using the same technique.

\begin{theorem}\label{th-hyper}
The set of minimal \scc{}s of a hypergraph $\HH = (N,E)$
can be computed in time $O(\size(\HH) \times \alpha(\card{N}))$, where
$\alpha$ denotes the inverse of the Ackermann function.
\end{theorem}

\section{The tropical double description method}\label{sec:ddm}

\begin{figure}[t]
\begin{center}
\begin{minipage}[b]{0.52\textwidth}
\vspace{0ex}
\begin{center}
\begin{tikzpicture}
[scale=0.65,>=triangle 45,
vtx/.style={circle,text=mygreen,fill=mygreen,inner sep=0pt,minimum size=1.25ex},
vtxlbl/.style={label distance=0pt},
ray/.style={rectangle,text=myred,fill=myred,inner sep=0pt,minimum size=1.15ex},
raylbl/.style={label distance=0pt},
new/.style={diamond,text=myblue,fill=myblue,inner sep=0pt,minimum size=1.5ex},
newlbl/.style={label distance=0pt}]
\equilateral{7}{90};

\barycenter{g1}{\expo{0}}{0}{\expo{0}};
\barycenter{g12}{\expo{0}}{\expo{0}}{\expo{0}};
\barycenter{g2}{\expo{2}}{\expo{2}}{\expo{0}};
\barycenter{g3}{0}{\expo{0}}{0};
\barycenter{g4}{\expo{-2}}{\expo{1}}{\expo{0}};
\barycenter{g41}{\expo{0}}{\expo{1}}{\expo{0}};

\barycenter{c1}{\expo{0}}{0}{0};
\barycenter{c2}{0}{\expo{2.5}}{\expo{0}};
\barycenter{c3}{0}{0}{\expo{0}};

\filldraw[lightgray,draw=black,very thick] (g1) -- (g12) -- (g2) -- (g3) -- (g4) -- (g41) -- cycle;

\filldraw[gray!40,draw=none,fill opacity=0.3] (c1) -- (c2) -- (c3) -- cycle;
\draw[gray, very thick] (c1) -- (c2) -- (c3);

\node[ray] at (g3) [label={[raylbl]above right:$\vect{g}^0$}] {};
\node[vtx] at (g4) [label={[vtxlbl,label distance=-0.5ex]above left:$\vect{g}^1$}] {};
\node[vtx] at (g2) [label={[vtxlbl]left:$\vect{g}^2$}] {};
\node[vtx] at (g1) [label={[vtxlbl]above right:$\vect{g}^3$}] {};

\node[new] at (intersection cs:
  first line={(c1) -- (c2)},
  second line={(g3) -- (g2)}) [label={[newlbl,label distance=-0.5ex]below:$\vect{h}^{2,0}$}] {};

\node[new] at (intersection cs:
  first line={(c1) -- (c2)},
  second line={(g3) -- (g1)}) [label={[newlbl,label distance=-1ex]above left:$\vect{h}^{3,0}$}] {};

\node[new] at (intersection cs:
  first line={(c1) -- (c2)},
  second line={(g3) -- (g4)}) [label={[newlbl,label distance=-0.25ex]above right:$\vect{h}^{1,0}$}] {};
\end{tikzpicture}
\end{center}
%
\caption{Intersecting a cone with a halfspace}\label{fig:tropical_poly2}
\end{minipage}
\hfill
\begin{minipage}[b]{0.47\textwidth}
\vspace{0ex}
\hskip-3mm\includegraphics[scale=0.47]{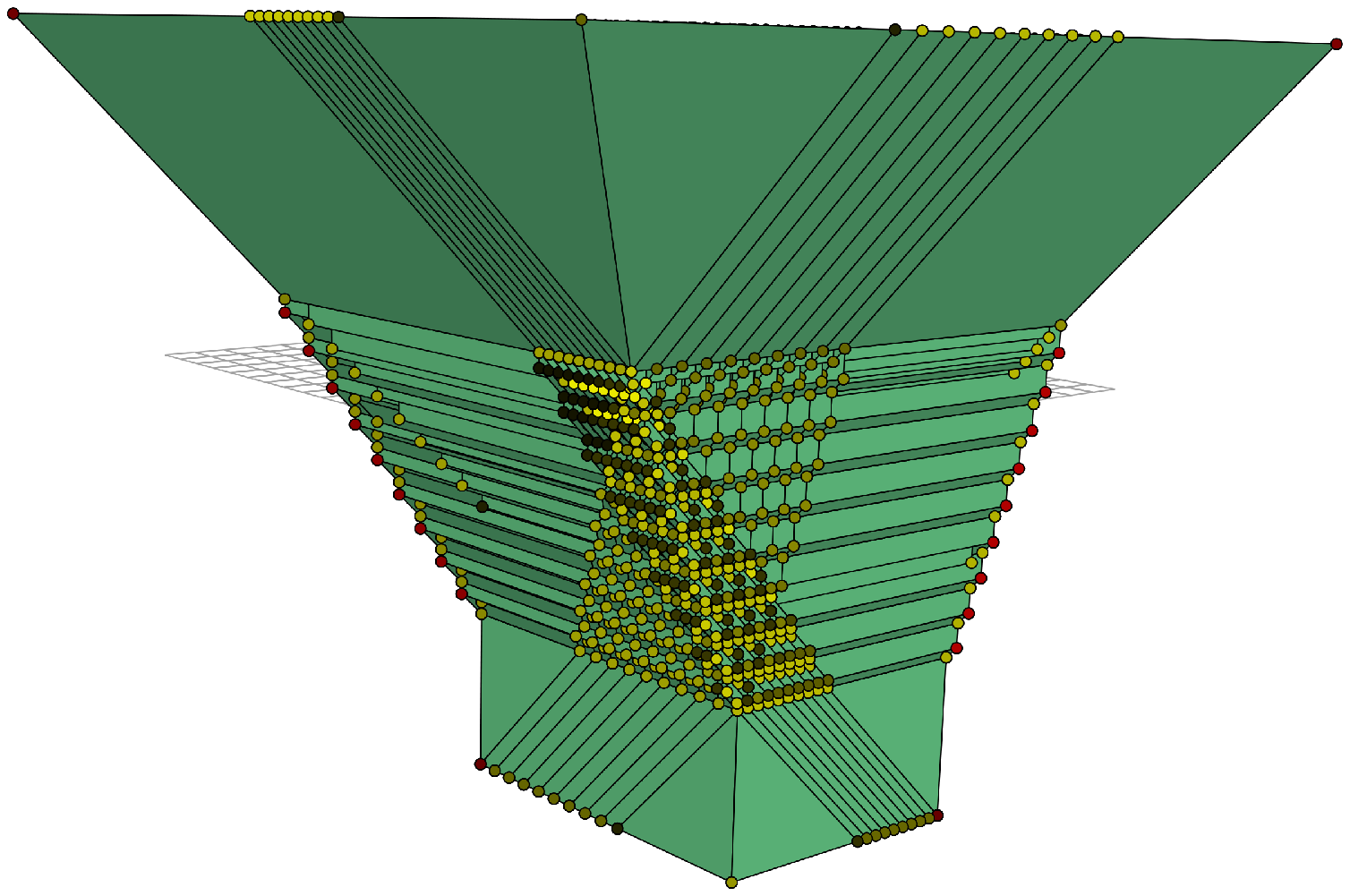}
\caption{Intersecting $10$ affine hyperplanes in dimension $3$}\label{fig:tropical_poly3}
\end{minipage}
\hfill
\end{center}
\end{figure}

Our algorithm is based on a successive elimination of inequalities. Given a polyhedral cone $\CC$ defined by a system of $n$ constraints, the algorithm computes by induction on $k$ ($0 \leq k \leq n$) a generating set $G_k$ of the intermediate cone defined by the first $k$ constraints. Then $G_n$ forms a generating set of the cone $\CC$. 
Passing from the set $G_k$ to the set $G_{k+1}$ relies on a result which, given a polyhedral cone $\KK$ and a tropical halfspace $\HH = \{\, \vect{x} \mid \vect{a} \vect{x} \leq \vect{b} \vect{x} \,\}$, allows to build a generating set $G'$ of $\KK \cap \HH$ from a generating set $G$ of $\KK$:
\begin{theorem}\label{th:ddm}
Let $\KK$ be a polyhedral cone generated by a set $G \subset \maxplus^d$, and $\HH = \{\, \vect{x} \mid \vect{a} \vect{x} \leq \vect{b} \vect{x} \,\}$ a tropical halfspace ($\vect{a}, \vect{b} \in \maxplus^{1 \times d}$). Then the polyhedral cone $\KK \cap \HH$ is generated by the set
\(
\{\,\vect{g} \in G \mid \vect{a} \vect{g} \leq \vect{b} \vect{g}\,\} \ \cup \  \{\, (\vect{a} \vect{h}) \vect{g} \mpplus (\vect{b} \vect{g}) \vect{h} \mid \vect{g},\vect{h} \in G \text{, } \vect{a} \vect{g} \leq \vect{b} \vect{g} \text{, and }
\vect{a} \vect{h} > \vect{b} \vect{h} \,\} .
\)
\end{theorem}
For instance, consider the cone defined in Fig.~\ref{fig:tropical_poly} and the constraint $\vect{x}_2 \leq \vect{x}_3 + 2.5$ (depicted in semi-transparent gray in Fig.~\ref{fig:tropical_poly2}). The three generators $\vect{g}^1$, $\vect{g}^2$, and $\vect{g}^3$ satisfy the constraint, while $\vect{g}^0$ does not. Their combinations are the elements $\vect{h}^{1,0}$, $\vect{h}^{2,0}$, and $\vect{h}^{3,0}$ respectively. 
The resulting algorithm is given in Figure~\ref{fig:computeextreme}. As in the classical case, this inductive approach produces redundant generators, hence, the heart of the algorithm is the extremality test in Line~10. We use
here the hypergraph characterization (Theorems~\ref{th:minimality} and~\ref{th-hyper}).

\begin{figure}[ht]
\vskip-3ex
\begin{scriptsize}
\begin{algorithmic}[1]
\Procedure {ComputeExtreme}{$A, B, n$} \Lcomment{$A, B \in \maxplus^{n \times d}$}
  \If{$n = 0$} \Lcomment{Base case} \label{ce:base}
    \State \Return the tropical canonical basis $(\vect{\epsilon}_i)_{1 \leq i \leq d}$
  \Else{} \Lcomment{Inductive case} \label{ce:inductive}
    \State split $A \vect{x} \leq B \vect{x}$ into $C \vect{x} \leq D \vect{x}$ and $\vect{a} \vect{x} \leq \vect{b} \vect{x}$, 
	   with $C, D \in \maxplus^{(n-1) \times d}$ and $\vect{a}, \vect{b} \in \maxplus^{1 \times d}$
    \State $G \gets \Call{ComputeExtreme}{C, D, n-1}$\label{ce:call}
    \State $G^\leq \gets \{\, \vect{g}^i \in G \mid \vect{a} \vect{g}^i \leq \vect{b} \vect{g}^i \, \}$, $G^> \gets \{\, \vect{g}^j \in G \mid \vect{a} \vect{g}^j > \vect{b} \vect{g}^j \, \}$, $H \gets G^\leq$ \label{ce:begin_elementary}
    \ForAll{$\vect{g}^i \in G^\leq$ and $\vect{g}^j \in G^>$}\label{ce:begin_loop}
      \State $\vect{h} \gets (\vect{a}\vect{g}^j)\vect{g}^i \mpplus (\vect{b} \vect{g}^i) \vect{g}^j$
      \If{$\vect{h}$ is extreme in $\{\, \vect{x} \mid A \vect{x} \leq B \vect{x}\,\}$}\label{ce:test}
	\State append $\kappa \vect{h}$ to $H$, where $\kappa$ is the opposite of the first non-$\mpzero$ coefficient of $\vect{h}$
      \EndIf
    \EndFor\label{ce:end_loop} \label{ce:end_elementary}
  \EndIf
  \State \Return $H$
\EndProcedure
\end{algorithmic}
\end{scriptsize}
\caption{Our main algorithm computing the extreme rays of tropical cones} \label{fig:computeextreme}
\end{figure}



\paragraph{\textit{Complexity analysis.}} The complexity of the elementary step of \Call{ComputeExtreme}{}, \ie{} the computation of the elements provided by Th.~\ref{th:ddm} and the elimination of non-extreme ones (Lines~\lineref{ce:begin_elementary} to~\lineref{ce:end_elementary}), can be precisely characterized to $O(n d \alpha(d) \card{G}^2)$, where $G$ is the generating set of the last intermediate cone. By comparison, for classical polyhedra, the same step in the refined double description method by Fukuda and Prodon~\cite{FukudaProdon96} takes
a time $O(n \card{G}^3)$. Note that $\card{G}$ can take values much larger
that $d$.

The overall complexity of the algorithm \Call{ComputeExtreme}{} depends on the size of the sets returned in the intermediate steps. 
In classical geometry, the upper bound theorem of McMullen~\cite{mcmullen70} shows that the maximal number of extreme points of a convex polytope in $\mathbb{R}^d$ defined by $n$ inequality constraints is equal to 
\[ \upperbound{n}{d} \defin \mychoose{n-\lfloor (d+1)/2\rfloor}{n-d} + \mychoose{n-\lfloor (d+2)/2\rfloor}{n-d}
\enspace .
\] 
The polars of the \emph{cyclic polytopes} (see~\cite{ziegler98}) are known to reach this bound. Allamigeon, Gaubert, and Katz~\cite{AGK09} showed that a similar bound is valid in the tropical setting.
\begin{theorem}[{\cite{AGK09}}]\label{th-upperbound}
The number of extreme rays of a tropical cone in $(\rbar)^d$
defined as the intersection of $n$ tropical half-spaces cannot exceed $U(n+d,d-1)=O((n+d)^{\lfloor (d-1)/2\rfloor }$. 
\end{theorem}
The bound is asymptotically tight for a fixed $n$,
as $d$ tends to infinity, being approached by a tropical generalization of the
(polar of) the cyclic polytope~\cite{AGK09}. The bound is believed not to be tight for a fixed $d$, 
as $n$ tends to infinity.
Finding the optimal bound is an open problem. 
By combining Theorem~\ref{th-upperbound}, Theorem~\ref{th-hyper}, and Theorem~\ref{th:minimality}, we readily get the following complexity result,
showing that the execution time is smaller in the tropical case
than in the classical case, even with the refinements of~\cite{FukudaProdon96}.
\begin{proposition}\label{wc}
The hypergraph implementation of the tropical double description method
returns the set of extreme rays
of a polyhedral cone defined by $n$ inequalities in dimension $d$
in time $O(n^2d\alpha(d)G_{\max}^2)$, where $G_{\max}$ is the
maximal number of extreme rays of a cone defined by a subsystem of inequalities
taken from $A\vect{x}\leq B\vect{x}$. 
In particular, the time can be bounded by $O(n^2 d \alpha(d) (n+d)^{d-1})$.
\end{proposition}

\paragraph{\textit{Alternative approaches.}}

The existing approachs discussed in the introduction have a structure which is similar to \Call{ComputeExtreme}{}. However, their implementation in the Maxplus toolbox of {\sc Scilab}~\cite{toolbox} and in our previous work~\cite{AGG08} relies on a much less efficient elimination of redundant generators. Its principle is the following: an element $\vect{h}$ is extreme in the cone generated by a given set $H$ if and only if $\vect{h}$ can not be expressed as the tropical linear combination of the elements of $H$ which are not proportional to it. This property can be checked in $O(d \times \card{H})$ time using residuation (see~\cite{BSS} for algorithmic details). In the context of our algorithm, the worst case complexity of the redundandy test is therefore $O(d\card{G}^2)$, where $G$ is the set of the extreme rays of the last intermediary cone. This is much worse that our method in $O(n d \alpha(d))$ based on directed hypergraphs, since the cardinality of the set $G$ may be exponential in $d$ (see Theorem~\ref{th-upperbound}). This is also confirmed by our experiments (see below).

%
%



We next sketch a different method relying on arrangement of tropical
hyperplanes (arrangements of classical hyperplanes
yield naive bounds). Indeed, Theorem~\ref{th:minimality}
implies that every extreme ray belongs to the intersection
of $d-1$ tropical hyperplanes, 
obtained by saturating $d-1$ inequalities among the $n+d$ taken
from $A\vect{x}\leq B\vect{x}$ and $\vect{x}_i\geq -\infty$, for $i\in [d]$.
The max-plus Cramer theorem (see~\cite{AGG08b} and the references therein)
implies that for generic values of the matrices $A,B$, every choice of $d-1$
saturated inequalities yields at most one candidate to be
an extreme ray, which can be computed
in $O(d^3)$ time. This yields a list of $O((n+d)^{d-1})$ candidates,
from which the extreme rays can be extracted by using the present hypergraph characterization (Theorems~\ref{th:minimality} and~\ref{th-hyper}), leading
to a $O((nd\alpha(d)+d^3)(n+d)^{d-1})$ execution time, which is better than the one of Proposition~\ref{wc} by a factor $n/\alpha(d)$ when $n\gg d$.
However, the resulting algorithm is of little practical use, since 
the worst case execution time is essentially always achieved, whereas
the double description method takes advantage of the fact that $G_{\max}$
is in general much smaller than the upper bound of Theorem~\ref{th-upperbound}
(which is probably not optimal in the case $n\gg d$).

A third approach, along the lines of~\cite{DS,joswig-2008}, would consist
in representing tropical polyhedra by polyhedral complexes in the usual sense.
However, an inconvenient of polyhedral complexes is that their number
of vertices (called ``pseudo-vertices'' to avoid ambiguities) is exponential
in the number of extreme rays~\cite{DS}. Hence, the representations used here
are more concise. 
This is illustrated in Figure~\ref{fig:tropical_poly3} (generated
using {\sc Polymake}), which shows an intersection of 10 signed tropical hyperplanes, corresponding
to the ``natural'' pattern studied in~\cite{AGK09}. There are only 24 extreme rays, but 1215 pseudo-vertices.

\paragraph{\textit{Experiments.}}

\begin{table}[t]
\vskip-5ex
\caption{Benchmarks on a single core of a $3 \, \, \mathrm{GHz}$ Intel Xeon with $3 \, \, \mathrm{Gb}$ RAM}\label{tab:experiments}
\begin{center}
\begin{tiny}
\begin{tabular}{|c|c|c|c|c|c|c|c|}
\hline
 & $d$ & $n$ & \# final & \# inter. & $T$ (s) & $T'$ (s) & $T/T'$\\
\hline
rnd$100$ & $12$ & $15$ & $32$ & $59$ & $0.24$ & $6.72$ & $0.035$ \\
rnd$100$ & $15$ & $10$ & $555$ & $292$ & $2.87$ & $321.78$ & $8.9 \cdot 10^{-3}$ \\
rnd$100$ & $15$ & $18$ & $152$ & $211$ & $6.26$ & $899.21$ & $7.0 \cdot 10^{-3}$ \\
rnd$30$ & $17$ & $10$ & $1484$ & $627$ & $15.2$ & $4667.9$ & $3.3 \cdot 10^{-3}$ \\
rnd$10$ & $20$ & $8$ & $5153$ & $1273$ & $49.8$ & $50941.9$ & $9.7 \cdot 10^{-4}$ \\
rnd$10$ & $25$ & $5$ & $3999$ & $808$ & $9.9$ & $12177.0$ & $8.1 \cdot 10^{-4}$ \\
rnd$10$ & $25$ & $10$ & $32699$ & $6670$ & $3015.7$ & --- & --- \\
cyclic & $10$ & $20$ & $3296$ & $887$ & $25.8$ & $4957.1$ & $5.2 \cdot 10^{-3}$ \\
cyclic & $15$ & $7$ & $2640$ & $740$ & $8.1$ & $1672.2$ & $5.2 \cdot 10^{-3}$ \\
cyclic & $17$ & $8$ & $4895$ & $1589$ & $44.8$ & $25861.1$ & $1.7 \cdot 10^{-3}$ \\
cyclic & $20$ & $8$ & $28028$ & $5101$ & $690$ & $45 \text{ days}$ & $1.8 \cdot 10^{-4}$ \\
cyclic & $25$ & $5$ & $25025$ & $1983$ & $62.6$ & $8 \text{ days}$ & $9.1 \cdot 10^{-5}$ \\
cyclic & $30$ & $5$ & $61880$ & $3804$ & $261$ & --- & --- \\
cyclic & $35$ & $5$ & $155040$ & $7695$ & $1232.6$ & --- & --- \\
\hline
\end{tabular}
\vskip0.5ex
\begin{tabular}{|c|c|c|c|c|c|c|}
\hline
& \# var & \# lines & $T$ (s) & $T'$ (s) & $T/T'$\\
\hline
\text{oddeven8} & $17$ & $118$ & $7.6$ & $152.1$ & $0.050$ \\
\text{oddeven9} & $19$ & $214$ & $128.0$ & $22101.2$ & $5.8 \cdot 10^{-3}$\\
\text{oddeven10} & $21$ & $240$ & $1049.0$ & --- & --- \\
\hline
\end{tabular}
\end{tiny}
\end{center}
\end{table}

Allamigeon has implemented Algorithm~\ComputeExtremal{} in OCaml,
as part of the ``Tropical polyhedral library'' ({\tt TPLib}), \href{http://penjili.org/tplib.html}{http://penjili.org/tplib.html}. Table~\ref{tab:experiments} reports some experiments for different classes of tropical cones: samples formed by several cones chosen randomly (referred to as rnd$x$ where $x$ is the size of the sample), and signed cyclic cones which are known to have a very large number of extreme elements~\cite{AGK09}. 
The successive columns respectively report the dimension $d$, the number of constraints $n$, the size of the final set of extreme rays, the mean size of the intermediary sets, and the execution time $T$ (for samples of ``random'' cones, we give average results). 

The result provided by~\ComputeExtremal{} does not depend on the order of the inequalities in the initial system. This order may impact the size of the intermediary sets and subsequently the execution time. In our experiments, inequalities are dynamically ordered during the execution: at each step of the induction, the inequality $\vect{a} \vect{x} \leq \vect{b} \vect{x}$ is chosen so as to minimize the number of combinations $(\vect{a}\vect{g}^j)\vect{g}^i \mpplus (\vect{b} \vect{g}^i) \vect{g}^j$. 
This strategy reports better results than without ordering.

We compare our algorithm with a variant using the alternative extremality criterion which is discussed in Sect.~\ref{sec:ddm} and used in the other existing implementations~\cite{toolbox,AGG08}. Its execution time $T'$ is given in the seventh column. The ratio $T/T'$ shows that our algorithm brings a huge breakthrough in terms of execution time. When the number of extreme rays is of order of $10^4$, the second algorithm needs several days to terminate. Therefore, the comparison could not be made in practice for some cases.

Table~\ref{tab:experiments} also reports some benchmarks from applications to static analysis. The experiments oddeven$i$ correspond to the static analysis of the odd-even sorting algorithm of $i$ elements. It is a sort of worst case for our analysis. The number of variables and lines in each program is given in the first columns. The analyzer automatically shows that the sorting program returns an array in which the last (resp.\ first) element is the maximum (minimum) of the array given as input. It clearly benefits from the improvements of \Call{ComputeExtreme}{}, as shown by the ratio with the execution time $T'$ of the previous implementation of the static analyzer~\cite{AGG08}.
\bibliographystyle{alpha}
\bibliography{tropical,biblio}
\end{document}